\documentclass[twocolumn,showpacs,preprintnumbers]{revtex4}
\usepackage{amssymb}

\usepackage{graphicx}
\usepackage{bm}

\begin{document}

\title{Non Abelian Berry Phase in Noncommutative Quantum Mechanics}
\author{Alain B\'{e}rard and Herv\'{e} Mohrbach}

\affiliation{Laboratoire de Physique Mol\'eculaire et des Collisions,
Institut de Physique, Technop\^ole 2000, 57078 Metz, France}

\date{\today}

\begin{abstract}
We consider the adiabatic evolution of the Dirac equation in order to
compute its Berry curvature in momentum space. It is found that the position
operator acquires an anomalous contribution due to the non Abelian Berry
gauge connection making the quantum mechanical algebra noncommutative. A
generalization to any known spinning particles is possible by using the
Bargmann-Wigner equation of motions. The non-commutativity of the
coordinates is responsible for the topological spin transport of spinning
particles similarly to the spin Hall effect in spintronic physics or the
Magnus effect in optics. As an application we predict new dynamics for
non-relativistic particles in an electric field and for photons in a
gravitational field.
\end{abstract}

\maketitle


\affiliation{Laboratoire de Physique Mol\'eculaire et des Collisions,
Institut de Physique, Technop\^ole 2000, 57078 Metz, France}


\section{Introduction}

Recently, Quantum Mechanics involving non-commutative space time coordinates
has led to numerous works. The assumption that the coordinate
operators do not commute was originally introduced by Snyder \cite{SNYDER}
as a short distance regularization to resolve the problem of
infinite energies in Quantum Field Theory. This idea became popular
when Connes \cite{CONNES} analyzed Yang Mills theories on non-commutative
space. More recently a correspondence between a non-commutative gauge theory
and a conventional gauge theory was introduced by Seiberg and Witten \cite{SEIBERG-WITTEN}. 
Non-commutative gauge theories are also naturally related
to string and M-theory \cite{KONECHNY} and to Galilean symmetry in the plane 
\cite{HORVATHY1}.

Applications of non-commutative theories were also found in condensed matter
physics, for instance in the Quantum Hall effect \cite{BELLISSARD} and the
non-commutative Landau problem \cite{GAMBOA}. Recently, it was found that a
non-commutative geometry also underlies the semiclassical dynamics of
electrons in semi-conductors \cite{MURAKAMI}. In this case, the
non-commutativity property of the coordinates originates from the presence
of a Berry phase which by changing drastically the equations of motion, 
induces a purely topological and dissipationless spin current. Other
equations of motion including a contribution of a Berry phase were also
recently found for the propagation of Bloch electrons \cite{NIU}.

In this paper we show that a non-commutative geometry underlies the
algebraic structure of all known spinning particles. In the Foldy-Wouthuysen
representation of the Dirac equation, the position operator acquires a
spin-orbit contribution which turns out to be a gauge potential (Berry
connection). It is important to mention that anomalous contributions to the
position operator were already found some time ago in different contexts,
for instance in the Bloch representation of electrons in a periodic
potential \cite{LANDAU} and for electrons in narrow-gap semi-conductors
(where the spin-orbit term is called a Yafet term \cite{YAFET}). The common
feature in all these cases is that an anomalous contribution to the position
operator stems from the representation where the kinetic energy is diagonal
(FW or Bloch representation). When interband transitions (adiabatic motion)
are neglected the algebraic structure of the coordinates becomes
non-commutative.

Then, after having determined the new position operator for spinning
particles we propose to explore its consequences at the level of
semi-classical equations of motion in several physical situations. In
particular, our approach provides a new interpretation of the Magnus effect,
which was observed experimentally in optics. But first of all, we recall
some previous results we derived in the framework of non-commutative Quantum
mechanics from symmetry arguments only \cite{NOUS}.

\section{Monopole in momentum space}

In non-commutative Quantum mechanics an antisymmetric parameter $\theta ^{ij}
$ usually taken to be constant \cite{DERIGLAZOV} is introduced in the
commutation relation of the coordinates in the space manifold 
\begin{equation}
\left[ x^{i},x^{j}\right] =i\hbar \theta ^{ij}.
\end{equation}
In a recent paper \cite{NOUS} we generalized the quantum mechanics in
noncommutative geometry by considering a quantum particle whose coordinates
satisfy the deformed Heisenberg algebra 
\begin{equation}
\left[ x^{i},x^{j}\right] =i\hbar \theta ^{ij}(\mathbf{x},\mathbf{p}),
\end{equation}
\begin{equation}
\lbrack x^{i},p^{j}]=i\hbar \delta ^{ij},\text{ and }[p^{i},p^{j}]=0.
\end{equation}
From Jacobi identity  
\begin{equation}
\lbrack p^{i},\left[ x^{j},x^{k}\right]
]+[x^{j},[x^{k},p^{i}]]+[x^{k},[p^{i},x^{j}]]=0,
\end{equation}
we deduced the important property that the $\theta $ field is only momentum
dependent. An important consequence of the non-commutativity between the
coordinates is that neither the position operator does satisfy the usual law 
$[x^{i},L^{j}]=i\hbar \varepsilon ^{ijk}x_{k}$, nor the angular momenta
satisfy the standard $so(3)$ algebra $[L^{i},L^{j}]=i\hbar \varepsilon
^{ijk}L_{k}$. In fact we have 
\begin{equation}
\lbrack x^{i},L^{j}]=i\hbar \varepsilon ^{ijk}x_{k}+i\hbar \varepsilon
^{j}{}_{kl}p^{l}\theta ^{ik}(\mathbf{p}),
\end{equation}
and 
\begin{equation}
\lbrack L^{i},L^{j}]=i\hbar \varepsilon ^{ij}{}_{k}L^{k}+i\hbar \varepsilon
^{i}{}_{kl}\varepsilon ^{j}{}_{mn}p^{l}p^{n}\theta ^{km}(\mathbf{p}).
\end{equation}
To remedy this absence of generators of rotation in the noncommutative
geometry we had to introduce a generalized angular momentum 
\begin{equation}
\mathbf{J}=\mathbf{\ r}\wedge \mathbf{p}+\lambda \frac{\mathbf{p}}{p},
\end{equation}
that satisfies the $so(3)$ algebra. The position operator then transforms as
a vector under rotations i.e., $[x^{i},J^{j}]=i\hbar \varepsilon ^{ijk}x_{k}$%
. The presence of the dual Poincare momentum $\lambda \mathbf{p/}p$ leads to
a dual Dirac monopole in momentum space for the position algebra 
\begin{equation}
\left[ x^{i},x^{j}\right] =-i\hbar \lambda \varepsilon ^{ijk}\frac{p^{k}}{%
p^{3}}.  \label{xxnc}
\end{equation}
This result immediately implies that the coordinates of spinless particles
are commuting. Another consequence is the quantification of the helicity $%
\lambda =n\hbar /2$ that arises from the restoration of the translation
group of momentum that is broken by the monopole \cite{NOUS}\cite{JACKIW}.
Note also that other recent theoretical works concerning the anomalous Hall
effect in two-dimensional ferromagnets predicted a topological singularity
in the Brillouin zone \cite{ONODA}. In addition, in recent experiments a
monopole in the crystal momentum space was discovered and interpreted in
terms of an Abelian Berry curvature \cite{FANG}.

In quantum mechanics this construction may look formal because it is always
possible to introduce commuting coordinates with the transformation $\mathbf{%
R=r-p\wedge S/}p^{2}\mathbf{.}$ The angular momentum is then $\mathbf{%
J=R\wedge p+S}$ which satisfies the usual $so(3)$ algebra, whereas the
potential energy term in the Hamiltonian now contains spin-orbit
interactions $V$ $(\mathbf{R+p\wedge S/}p^{2})$. In fact, the inverse
procedure is usually more efficient: considering an Hamiltonian with a
particular spin-orbit interaction one can try to obtain a trivial
Hamiltonian with a dynamics due to the noncommutative coordinates algebra.
This procedure has been applied with success to the study of adiabatic
transport in semiconductor with spin-orbit couplings \cite{MURAKAMI} where
the particular dynamics of charges is governed by the 
commutation relation (\ref{xxnc}). The important point is to 
determine which one of the two position operators $\mathbf{r}$ or $\mathbf{R%
}$ gives rise to the real mean trajectory of the particle. In fact it is
well known that $\mathbf{R}$ does not have the genuine property of a
position operator for a relativistic particle. As we shall see this crucial
remark implies a new prediction concerning the non-relativistic limit of a
Dirac particle.

In particle physics it is by now well known that the non-commutativity of
the coordinates of massless particles is a fundamental property because the
position operator does not transform like a vector unless it satisfies
equation (\ref{xxnc}) and that $\theta ^{ij}(%
\mathbf{p})$ is the Berry curvature for a massless particle with a given
helicity $\lambda $ \cite{SKAGERSTAM}.

In this letter we present another point of view of the origin of the
monopole in high energy and condensed matter physics by considering the
adiabatic evolution of relativistic massive spinning particles. In
particular the computation of the Berry curvature of Dirac particles gives
rise to a noncommutative position operator that was already postulated by
Bacry \cite{BACRY} some time ago. A generalization to any spin is possible 
via the Bargmann-Wigner \cite{BARGMANN} equations of motion. By doing
that construction, we are brought to make a generalization of noncommutative
algebra by considering a $\theta $ field which is momentum as well as spin
dependent. The associated connection is then non Abelian but becomes Abelian
in the limit of vanishing mass leading to a monopole configuration for the
Berry curvature. In this respect our approach is different from \cite{SKAGERSTAM} 
because the description of the photons is obtained by taking
the zero mass limit of the massive representation of a spin one particle.

\section{The Foldy-Wouthuysen representation}

The Dirac's Hamiltonian for a relativistic particle of mass $m$ has the form 
\[
\hat{H}=\mathbf{\alpha .p}+\beta m+\hat{V}\left( \mathbf{R}\right) ,
\]
where $\hat{V}$ is an operator that acts only on the orbital degrees of
freedom. Using the Foldy-Wouthuysen unitary transformation 
\[
U(\mathbf{p})=\frac{E_{p}+mc^{2}+c\beta \mathbf{\alpha .p}}{\sqrt{%
2E_{p}\left( E_{p}+mc^{2}\right) }},
\]
with $E_{p}=\sqrt{p^{2}c^{2}+m^{2}c^{4}}$, we obtain the following
transformed Hamiltonian 
\[
U(\mathbf{p})\hat{H}U(\mathbf{p})^{+}=E_{p}\beta +U(\mathbf{p})\hat{V}%
(i\hbar \partial _{\mathbf{p}})U(\mathbf{p})^{+}.
\]
The kinetic energy is now diagonal whereas the potential term becomes $\hat{V%
}(\mathbf{D})$ with the covariant derivative defined by $\mathbf{\ D=}i\hbar
\partial _{\mathbf{p}}+\mathbf{A}$, and with the gauge potential $\mathbf{A}%
=i\hbar U(\mathbf{p})\partial _{\mathbf{p}}U(\mathbf{p})^{+}$, which reads 
\begin{equation}
\mathbf{A}=\frac{\hbar c\left( ic^{2}\mathbf{p}(\mathbf{\alpha .p})\beta
+i\beta \left( E_{p}+mc^{2}\right) E_{p}\mathbf{\alpha -}cE_{p}\mathbf{\
\Sigma \wedge p}\right) }{2E_{p}^{2}\left( E_{p}+mc^{2}\right) },
\end{equation}
where $\mathbf{\Sigma }=1\otimes \mathbf{\sigma }$, is a $\left( 4\times
4\right) $ matrix. We consider the adiabatic approximation by identifying
the momentum degree of freedom as slow and the spin degree of freedom as
fast, similarly to the nuclear configuration in adiabatic treatment of
molecular problems, which allows us to neglect the interband transition. We
then keep only the block diagonal matrix element in the gauge potential and
project on the subspace of positive energy. This projection cancels the
zitterbewegung which corresponds to an oscillatory motion around the mean
position of the particle that mixes the positive and negative energies. In
this way we obtain a non trivial gauge connection allowing us to define a
new position operator $\mathbf{r}$ for this particle 
\begin{equation}
\mathbf{r=}i\hslash \partial _{\mathbf{p}}+\frac{c^{2}\hslash \left( \mathbf{%
\ p}\wedge \mathbf{\sigma }\right) }{2E_{p}\left( E_{p}+mc^{2}\right) }\text{%
\textbf{\ ,}}  \label{r}
\end{equation}
which is a $\left( 2\times 2\right) $ matrix. The position operator (\ref{r}%
) is not new, as it was postulated by H. Bacry \cite{BACRY}. By considering
the irreducible representation of the Poincare group, this author proposed
to adopt a general position operator for free massive or massless particles
with any spin. In our approach which is easily generalizable to any known
spin (see formula (\ref{rs})) the anomalous part of the position operator
arises from an adiabatic process of an interacting system and as we will now
see is related to the Berry connection. For a different work with operator
valued position connected to the spin-degree of freedom see \cite{GHOSH2}.
Zitterbewegung-free noncommutative coordinates were also introduced for
massless particles with rigidity and in the context of anyons \cite{PLYUSHCHAY}.

It is straightforward to prove that the anomalous part of the position
operator can be interpreted as a Berry connection in momentum space which,
by definition is the $(4\times 4)$ matrix 
\begin{equation}
\mathbf{A}_{\alpha \beta }(\mathbf{p})=i\hbar <\alpha \mathbf{p}+\mid
\partial _{\mathbf{p}}\mid \beta \mathbf{p}+>
\end{equation}
where $\mid \alpha $ $\mathbf{p}+>$ is an eigenvector of the free Dirac
equation of positive energy. The Berry connection can also be written as 
\begin{equation}
\mathbf{A}_{\alpha \beta }(\mathbf{p})=i\hbar <\phi _{\alpha }\mid U\partial
_{\mathbf{p}}U^{+}\mid \phi _{\beta }>,
\end{equation}
in terms of the canonical base vectors $\mid \phi _{\alpha }>=\left( 
\begin{array}{llll}
1 & 0 & 0 & 0
\end{array}
\right) $ and $\mid \phi _{\beta }>=\left( 
\begin{array}{llll}
0 & 1 & 0 & 0
\end{array}
\right) $. With the non zero element belonging only to the
positive subspace, we can define the Berry connection by considering a $%
2\times 2$ matrix 
\begin{equation}
\mathbf{\ A}(\mathbf{p})=i\hbar \mathcal{P}(U\partial _{\mathbf{p}}U^{+}),
\end{equation}
where $\mathcal{P}$ is a projector on the positive energy subspace. In this
context the $\theta $ field we postulated in \cite{NOUS} emerges naturally
as a consequence of the adiabatic motion of a Dirac particle and corresponds
to a non-Abelian gauge curvature satisfying the relation 
\begin{equation}
\theta ^{ij}(\mathbf{p,\sigma })=\partial _{p^{i}}A^{j}-\partial
_{p^{j}}A^{i}+\left[ A^{i},A^{j}\right] .
\end{equation}
The commutation relations between the coordinates are then 
\begin{equation}
\left[ x^{i},x^{j}\right] =i\hslash \theta ^{ij}(\mathbf{p},\mathbf{\sigma }%
)=-i\hbar ^{2}\varepsilon _{ijk}\frac{c^{4}}{2E_{p}^{3}}\left( m\sigma ^{k}+%
\frac{p^{k}(\mathbf{p}.\mathbf{\sigma )}}{E_{p}+mc^{2}}\right).  \label{nc}
\end{equation}
This relation has very important consequences as it implies the
nonlocalizability of the spinning particles. This is an intrinsic property
and is not related to the creation of a pair during the measurement process
(for a detailed discussion of this important point see \cite{BACRY})

To generalize the construction of the position operator for a particle with
unspecified $n/2$ $(n>1)$ spin, we start with the Bargmann-Wigner equations 
\[
(\gamma _{\mu }^{(i)}\partial _{\mu }+m+\hat{V})\psi _{(a_{1}...a_{n})}=0\ \
\ \ \ \ \ \ \ (i=1,2...n),
\]
where $\psi _{(a_{1}...a_{n})}$ is a Bargmann-Wigner amplitude and $\gamma
^{(i)}$ are matrices acting on $a_{i}$. For each equation we have a
Hamiltonian 
\[
\hat{H}^{(i)}=\mathbf{\alpha }^{(i)}\mathbf{.p}+\beta m+\hat{V},
\]
then 
\begin{equation}
(\prod\limits_{j=1}^{n}U^{(j)}(\mathbf{p}))\hat{H^{(i)}}(\prod%
\limits_{j=1}^{n}U^{(j)}(\mathbf{p})^{+})=E_{p}\beta ^{(i)}+\hat{V}(\mathbf{D%
}),
\end{equation}
with $\mathbf{D=}i\hbar \partial _{\mathbf{p}}+\sum\limits_{i=1}^{n}\mathbf{A%
}^{(i)},$ and $\mathbf{A}^{(i)}=i\hbar U^{(i)}(\mathbf{p})\partial _{\mathbf{%
p}}U^{(i)}(\mathbf{p})^{+}$. Again by considering the adiabatic
approximation we deduce a general position operator $\mathbf{r}$ for
spinning particles 
\begin{equation}
\mathbf{r=}i\hslash \partial _{\mathbf{p}}+\frac{c^{2}\left( \mathbf{p}%
\wedge \mathbf{S}\right) }{E_{p}\left( E_{p}+mc^{2}\right) }\mathbf{,}
\label{rs}
\end{equation}
with $\mathbf{S=}\hslash \left( \mathbf{\sigma }^{(1)}+...+\mathbf{\sigma }%
^{(n)}\right) /2$. The generalization of (\ref{nc}) is then 
\begin{equation}
\left[ x^{i},x^{j}\right] =i\hslash \theta ^{ij}(\mathbf{p},\mathbf{S}%
)=-i\hbar \varepsilon _{ijk}\frac{c^{4}}{E_{p}^{3}}\left( mS^{k}+\frac{p^{k}(%
\mathbf{p}.\mathbf{S)}}{E_{p}+mc^{2}}\right) .
\end{equation}
For a massless particle we recover the relation $\mathbf{r=}i\hslash
\partial _{\mathbf{p}}+\mathbf{p}\wedge \mathbf{S/}p^{2}$, \ with the
commutation relation giving rise to the monopole $\left[ x^{i},x^{j}\right]
=i\hslash \theta ^{ij}(\mathbf{p})=-i\hbar \varepsilon _{ijk}\lambda \frac{%
p^{k}}{p^{3}}$. The monopole in momentum introduced in \cite{NOUS} in order
to construct genuine angular momenta has then a very simple physical
interpretation. It corresponds to the Berry curvature resulting from an
adiabatic process of a massless particle with helicity $\lambda $. For $%
\lambda =\pm 1$ we have the position operator of the photon, whose
non-commutativity property agrees with the weak localizability of the photon
which is certainly an experimental fact. It is not surprising that a
massless particle has a monopole Berry curvature as it is well known that
the band touching point acts as a monopole in momentum space \cite{BERRY}.
This is precisely the case for massless particles for which the positive and
negative energy band are degenerate in $p=0$. In our approach, the monopole
appears as a limiting case of a more general Non Abelian Berry curvature
arising from an adiabatic process of massive spinning particles.

The spin-orbit coupling term in (\ref{rs}) is a very small correction to the
usual operator in the particle physics context but it may be strongly
enhanced and observable in solid state physics because of the spin-orbit
effect being more pronounced than in the
vacuum. For instance in narrow gap semiconductors the equations of the band
theory are similar to the Dirac equation with the forbidden gap $E_{G}$
between the valence and conduction bands instead of the Dirac gap $2mc^{2}$ 
\cite{RASHBA2}.The monopole in momentum space predicted and observed in
semiconductors results from the limit of vanishing gap $E_{G}\rightarrow 0$
between the valence and conduction bands.

It is also interesting to consider the symmetry properties of the position
operator with respect to the group of spatial rotations. In terms of
commuting coordinates $\mathbf{R}$ the angular momentum is by definition $%
\mathbf{J}=\mathbf{R\wedge p}+\mathbf{S}$, whereas in terms of the
noncommuting coordinates the angular momentum reads $\mathbf{J}=\mathbf{\
r\wedge p}+\mathbf{M,}$ where 
\begin{equation}
\mathbf{M}=\mathbf{S-A\wedge p.}  \label{m}
\end{equation}
One can explicitly check that in terms of the non commuting coordinates the
relation $[x^{i},J^{j}]=i\hbar \varepsilon ^{ijk}x_{k}$ is satisfied, so $%
\mathbf{r}$ like $\mathbf{R}$ transforms as a vector under space rotations,
but $d\mathbf{R}/dt=c\mathbf{\alpha }$ which is physically unacceptable. For
a massless particle (\ref{m}) leads to the Poincar\'{e} momentum associated
to the monopole in momentum space deduced in \cite{NOUS}.

\section{Dynamical equations of motion}

Let us now look at some physical consequences of the non-commuting position
operator on the dynamics of a quantum particle in an arbitrary potential.
Due to the Berry phase in the definition of the position the equation of
motion should be changed. But to compute commutators like $\left[ x^{k},V(x)%
\right] $ one resorts to the semiclassical approximation $\left[ x^{k},V(x)%
\right] =i\hbar \partial _{l}V(x)\theta ^{kl}+O(\hbar ^{2})$ leading to new
equations of motion 
\begin{equation}
\stackrel{.}{\mathbf{r}}=\frac{\mathbf{p}}{E_{p}}-\stackrel{.}{\mathbf{p}}%
\mathbf{\wedge \theta }\text{, \qquad and \qquad }\stackrel{.}{\mathbf{p}}=-%
\mathbf{\nabla }V\mathbf{(r)}  \label{rp}
\end{equation}
with $\theta ^{i}=\varepsilon ^{ijk}\theta _{jk}/2$. While the
equation for the momentum is as usual, the one for the
velocity acquires a topological contribution due to the Berry phase. The latter 
is responsible for the relativistic topological spin
transport as in the context of semi-conductors where similar
non-relativistic equations \cite{MURAKAMI} lead to the spin Hall effect \cite{HIRSCH}.

\section{Applications}

\subsection{Non-relativistic Dirac particle in an electric potential}

As a particular application, consider the nonrelativistic limit of a charged
spinning Dirac particle in an electric potential $\hat{V}(\mathbf{r})$. In
the NR limit the Hamiltonian reads 
\begin{equation}
\widetilde{H}(\mathbf{R,p})\approx mc^{2}+\frac{p^{2}}{2m}+\hat{V}(\mathbf{R}%
)+\frac{e\hbar }{4m^{2}c^{2}}\mathbf{\sigma .}\left( \mathbf{\nabla \hat{V}(%
\mathbf{r})\wedge p}\right) ,  \label{H}
\end{equation}
which is a Pauli Hamiltonian with a spin-orbit term. As shown in \cite{MATHUR}, 
the nonrelativistic Berry phase $\theta ^{ij}=-\varepsilon
_{ijk}\sigma ^{k}/2mc^{2}$ results also from the Born-Oppenheimer
approximation of the Dirac equation which leads to the same non-relativistic
Hamiltonian. In the same paper, it was also proved that the adiabaticity
condition is satisfied for slowly varying potential such that $L>>\tilde{%
\lambda}$, where $L$ is the length scale over which $\hat{V}(\mathbf{r})$
varies and $\tilde{\lambda}$ is the de Broglie wave length of the particle.
From Hamiltonian (\ref{H}) we deduce the dynamics of the Galilean Schr\"{o}%
dinger position operator $\mathbf{R}$ 
\begin{equation}
\frac{dX^{i}}{dt}=\frac{p^{i}}{m}+\frac{e\hslash }{4m^{2}c^{2}}\varepsilon
^{ijk}\sigma _{j}\partial _{k}\hat{V}(\mathbf{r}),  \label{xnr1}
\end{equation}
whereas the non relativistic limit of $\left( \ref{rp}\right) $ leads to the
following velocity 
\begin{equation}
\frac{dx^{i}}{dt}=\frac{p^{i}}{m}+\frac{e\hslash }{2m^{2}c^{2}}\varepsilon
^{ijk}\sigma _{j}\partial _{k}\hat{V}(\mathbf{r}).  \label{xnr2}
\end{equation}
We then predict an enhancement of the spin-orbit coupling when the new
position operator is considered. One can appreciate the similarity between
this result and the Thomas precession as it offers another manifestation of
the difference between the Galilean limit $\left( \ref{xnr1}\right) $ and
the non-relativistic limit $\left( \ref{xnr2}\right) $.

\subsection{Rashba coupling}

Another interesting non relativistic situation concerns a parabolic quantum
well with an asymmetric confining potential $V(z)=m\omega^2z^2/2$ in a
normal electric field $E_{z}$ producing the structure inversion asymmetry.
By considering again the NR limit of the position operator (\ref{rs}), we
get a spin orbit coupling of the form $\frac{\hbar }{4m^{2}c^{2}}%
(eE_{z}+m\omega _{0}^{2}Z)\left( p_{x}\sigma_{y}-p_{y}\sigma_{x}\right)
+O\left( 1/m^{3}\right) $, which for strong confinement in the $(x,y)$ plane
is similar to the Rashba spin-orbit coupling well known in semi-conductor
spintronics \cite{RASHBA}. This effect is very small for non-relativistic
momenta, but as already said, it is greatly enhanced in semiconductors by a
factor of about $mc^2/E_G$.

\subsection{Ultrarelativistic particle in an electric field}

Another example of topological spin transport that we consider now arises in
the ultrarelativistic limit. In this limiting case $E_{p}\approx pc$ and the
equations of motion of the spinning particle in a constant electric field
are 
\begin{equation}
\frac{dx^{i}}{dt}=\frac{cp^{i}}{p}+\lambda e\varepsilon ^{ijk}\frac{p^{j}}{%
p^{3}}E_{k}.
\end{equation}
Taking the electric field in the $z$ direction and as initial conditions $%
p_{1}(0)=p_{3}(0)=0$ and $p_{2}(0)=p_{0}>>mc^{2}$, we obtain the coordinates
in the Heisenberg representation 
\begin{equation}
x\left( t\right) =\frac{\lambda }{p_{0}}\frac{eEt}{\left(
p_{0}^{2}+e^{2}E^{2}t^{2}\right) ^{1/2}},
\end{equation}
\begin{equation}
y(t)=\frac{p_{0}c}{eE}\arg \sinh \left( \frac{eEt}{p_{0}}\right) ,
\end{equation}
\begin{equation}
z(t)=\frac{c}{eE}\left[ \left( p_{0}^{2}+e^{2}E^{2}t^{2}\right) ^{1/2}-p_{0}%
\right] .
\end{equation}
We observe an unusual displacement in the $x$ direction perpendicular to the
electric field which depends on the value of the helicity. This topological
spin transport can be considered as a relativistic generalization of the
spin Hall effect discussed in \cite{MURAKAMI}. At large time the shift is of
the order of the particle wave length $\left| \Delta x\right| \sim \tilde{%
\lambda}$ which in ultrarelativistic limit is of the order of the Compton
wave length. This very small effect is obviously very difficult to observe
in particular due to the creation of pair particles during the measurement
process itself. Actually this effect has been already observed 
however only in the case of photons propagating in an inhomogeneous medium.

\subsection{Spin Hall effect of light}

Experimentally what we call a topological spin transport has been first
observed in the case of the photon propagation in an inhomogeneous medium 
\cite{ZELDOVICH}, where the right and left circular polarization propagate
along different trajectories in a wave guide (the transverse shift is
observable due to the multiple reflections), a phenomena interpreted quantum
mechanically as arising from the interaction between the orbital momentum
and the spin of the photon \cite{ZELDOVICH}. To interpret the experiments
these authors introduced a complicated phenomenological Hamiltonian leading
to generalized geometrical optic equation. Our approach provides a new
satisfactory interpretation as this effect, also called optical Magnus
effect, is now interpreted in terms of the non-commuting property of the
position operator containing the Berry phase. Note that the adiabaticity
conditions in this case are given in \cite{chao}. To illustrate our purpose
consider the Hamiltonian of a photon in an inhomogeneous medium $H=pc/n(r)$.
The equations of motion $\stackrel{\cdot }{x^{i}}=\frac{1}{i\hbar }\left[
x^{i},H\right] $ and $\stackrel{\cdot }{p^{i}}=\frac{1}{i\hbar }\left[
p^{i},H\right] $ in the semi-classical approximation leads to following
relations between velocities and momenta 
\begin{equation}
\frac{dx^{i}}{dt}=\frac{c}{n}\left( \frac{p^{i}}{p}+\frac{\lambda
\varepsilon ^{ijk}p_{k}}{p^{2}}\frac{\partial \ln n}{\partial x^{j}}\right) 
\label{nopt}
\end{equation}
which are similar to those introduced phenomenologically in \cite{ZELDOVICH}%
. However, here they are deduced rigorously from different physical
considerations. We readily observe that the Berry phase gives rise to an
''ultra-relativistic spin-Hall effect'' which in turn implies that the
velocity is no more equal to $c/n$. Note that similar equations are also
given in \cite{BLIOKH} where the optical Magnus effect is also interpreted
in terms of a monopole Berry curvature but in the context of geometric
optics.

\subsection{Photon in a static gravitational field}

Our theory is easily generalizable to the photon propagation in an
anisotropic medium, a situation which is simply mentioned in \cite{ZELDOVICH}
but could not be studied with their phenomenological approach. As a typical
anisotropic medium consider the photon propagation in a static gravitational
field whose metric $g^{ij}(x)$ is supposed to be time independent $\left(
g^{0i}=0\right) $ and having a Hamiltonian $H=c\left( -\frac{%
p_{i}g^{ij}(x)p_{j}}{g^{00}(x)}\right) ^{1/2}.$ In the semi-classical
approximation the equations of motion are 
\begin{equation}
\frac{dp_{k}}{dt}=\frac{c^{2}p_{i}p_{j}}{2H}\partial _{k}\left( \frac{%
g^{ij}(x)}{g^{00}(x)}\right)   \label{pt}
\end{equation}
and 
\begin{equation}
\frac{dx^{k}}{dt}=\frac{c\sqrt{g_{00}}g^{ki}p_{i}}{\sqrt{-g^{ij}p_{i}p_{j}}}+%
\frac{dp_{l}}{dt}\theta ^{kl}  \label{xt}
\end{equation}
For a static gravitational field the velocity is then 
\begin{equation}
v^{i}=\frac{c}{\sqrt{g_{00}}}\frac{dx^{i}}{dx^{0}}=c\frac{g^{ij}p_{j}}{\sqrt{%
-g^{ij}p_{i}p_{j}}}+\frac{1}{\sqrt{g_{00}}}\frac{dp_{l}}{dt}\theta ^{kl}
\end{equation}
with $x^{0}=ct$. Equations (\ref{pt}) and (\ref{xt}) are our new equations
for the semiclassical propagation of light which take into account the
non-commutative nature of the position operator, i.e the spin-orbit coupling
of the photon. The spinning nature of photon introduces a quantum Berry
phase, which affects the propagation of light in a static background
gravitational field  at the semi-classical level. This new fundamental
prediction will be studied in more detail in a future paper, but we can
already observe that the Berry phase implies a speed of light different from
the universal value $c$. This effect which is still very small could become
important for a photon being propagated in the gravitational field of a
black hole. This result goes in the same direction as recent works on the
possibility of a variable speed of light \cite{magueijo} but here this
variation has a physical origin.

\section{Conclusion}

In summary, we looked at the adiabatic evolution of the Dirac equation in
order to clarify the relation between monopole and Berry curvature in
momentum space. It was found that the position operator acquires naturally
an anomalous contribution due to a non Abelian Berry gauge connection making
the quantum mechanical algebra non-commutative. Using the Bargmann-Wigner
equation of motions we generalized our formalism to all known spinning
particles. The non-commutativity of the coordinates is responsible for the
topological spin transport of spinning particles similarly to the spin Hall
effect in spintronic physics or the optical Magnus effect in optics. In
particular we predict two new effects. One is an unusual spin-orbit
contribution of a non-relativistic particle in an external field. The other
one concerns the effect of the Berry phase on the propagation of light in a
static background gravitational field.


\begin{thebibliography}{99}
\bibitem{SNYDER}  H.\ Snyder, Phys. Rev.\textbf{71} (1947) 38;

\bibitem{CONNES}  A.\ Connes, ''Noncommutative Geometry'', Academic press,
San Diego (1994);

\bibitem{SEIBERG-WITTEN}  N.\ Seiberg, E.\ Witten, JHEP. \textbf{09} (1999)
32;

\bibitem{KONECHNY}   A.\ Konechny, A. Schwarz, Phys. Rep. \textbf{360}
(2002) 353.

\bibitem{HORVATHY1}  C. Duval and P. A. Horvathy Phys. lett. B. \textbf{479}
(2000) 284; J. Phys. A \textbf{34} (2001) 10097.

\bibitem{BELLISSARD}  J.\ Belissard, Lecture Notes in Physics \textbf{257}
(1986) 99; J.\ Belissard, A. Van Elst and H.\ Schulz-Baldes, cond-mat/
9301005.

\bibitem{GAMBOA}  J.\ Gamboa et al., Mod. Phys. Lett. A \textbf{16} (2001)
2075; P. A. Horvathy, Ann. Phys. \textbf{299} (2002) 128.

\bibitem{MURAKAMI}  S. Murakami, N. Nagaosa, S.C. Zhang, Science \textbf{301}
(2003) 1348.

\bibitem{NIU}  G. Sundaram, Q. Niu, Phys. Rev. B \textbf{59} (1999) 14915.

\bibitem{LANDAU}  E. M. Lifshitz and L. P. Pitaevskii, ''Statistical
Physics'', Vol \textbf{9}, Pergamon Press (1981).

\bibitem{YAFET}  Y. Yafet, ''Solid State Physics'', Vol \textbf{14},
Academic, New York, (1963)

\bibitem{NOUS}  A.\ B\'{e}rard, H.\ Mohrbach, Phys. Rev. D \textbf{69}
(2004) 127701; A. B\'{e}rard, Y. Grandati, H. Mohrbach, Phys. lett. A 
\textbf{254} (1999) 133; A. B\'{e}rard, J. Lages, H. Mohrbach, Eur. Phys. J.
C \textbf{35} \textbf{2004} (373).

\bibitem{DERIGLAZOV}  A. A. Deriglazov, JHEP \textbf{03}, (2003) 021; Phys.
Lett. B \textbf{555 } (2003) 83; M. Sheikh-Jabbari, Nucl. Phys. B \textbf{611%
} (2001) 383; M.\ Chaichan, M. Sheikh-Jabbari and A.\ Tureanu, Phys. Rev.
Lett., \textbf{86} (2001) 2716; S.\ Ghosh, Phys. Rev. D \textbf{66} (2002)
045031; J. Romero, J. Santiago and J. D. Vergara, Phys. Lett. A \textbf{310}%
, (2003) 9.

\bibitem{JACKIW}  R.\ Jackiw, Phys. Rev. Lett. \textbf{54} (1985) 159 and
hep-th /0212058.

\bibitem{ONODA}  M.\ Onoda, N.\ Nagasoa, J.\ Phys. Soc. Jpn. \textbf{71}
(2002) 19.

\bibitem{FANG}  Z.\ Fang et al., Science \textbf{302} (2003) 92.

\bibitem{SKAGERSTAM}  B.S. Skagerstam hep-th /9210054.

\bibitem{BACRY}  H.\ Bacry, ''Localizability and Space in Quantum Physics'',
Lecture Notes in Physics, Vol \textbf{308}, Heidelberg, Springer-Verlag,
(1988).

\bibitem{BARGMANN}  V.\ Bargmann, E.P. Wigner, Proc. Nat. Acd. Sci. \textbf{%
34} (1948) 211.

\bibitem{GHOSH2}  S.\ Ghosh, Phys. Lett. B \textbf{571} (2003) 97.

\bibitem{PLYUSHCHAY}  M.S. Plyushchay, Mod. Phys. Lett. A \textbf{4} (1989)
837; Phys. Lett. B \textbf{243} (1990) 383; J.L. Cortes, M.S. Plyushchay,
Int. J. Mod. Phys. A \textbf{11} (1996) 3331; P.A. Horvathy, M.S.
Plyushchay, JHEP \textbf{0206} (2002) 033; hep-th/0404137.

\bibitem{BERRY}  M.V.\ Berry, Proc. R.\ Soc. London A \textbf{392} (1984) 45.

\bibitem{RASHBA2}  E.I. Rashba, Physica E, \textbf{20} (2004) 189.

\bibitem{HIRSCH}  J.E.\ Hirsch, Phys. Rev. Lett. \textbf{83} (1999) 1834.

\bibitem{MATHUR}  H.\ Mathur, Phys. Rev. Lett. \textbf{67} (1991) 3325.

\bibitem{RASHBA}  Yu. Bychkov and E.I. Rashba, JETP Lett. \textbf{39} (1984)
78.

\bibitem{ZELDOVICH}  A.V. Dooghin et al., Phys. Rev. A \textbf{45} (1992)
8204; V.S. Liberman, B.Y.\ Zeldovich, Phys. Rev. A \textbf{46} (1992) 5199.

\bibitem{chao}  R.Y. Chiao, Y. Wu, Phys. Rev. Lett. \textbf{57} (1986) 933.

\bibitem{BLIOKH}  K.Y. Bliokh, Y.P. Bliokh, Phys. Lett A \textbf{333}
(2004), 181; Phys. Rev. E \textbf{70} (2004) 026605. M. Onoda, S. Murakami,
N. Nagasoa, Phys. Rev. Lett. \textbf{93} (2004) 083901.

\bibitem{magueijo}  J. Magueijo, Rep. Prog. Phys. \textbf{66} (2003) 2025.
\end{thebibliography}
\end{document}